# Sizing and Location Selection of Medium-Voltage Back-to-Back Converter for DER-Dominated Distribution Systems

Xiangqi Zhu, *Senior Member, IEEE*, Akanksha Singh, *Senior Member, IEEE*, Barry Mather, *Senior Member, IEEE*

*Abstract*— Medium-voltage back-to-back (MVB2B) converters can connect two distribution systems and quantifiably transfer power between them. This function can enable the MVB2B converter to exchange distributed energy resource (DER)-generated power between two systems and bring significant value to enhancing distribution system DER adoption. Our previous work analyzed and demonstrated the value the MVB2B converter can bring to DER integration. As continuous work, this paper presents a methodology that helps address the MVB2B converter sizing and location selection problem in distribution systems with high DER penetrations. The proposed methodology aims to address three critical problems for MVB2B converter implementation in the real world: 1) which distribution systems are better to be connected, 2) what converter size is appropriate for connecting the distribution systems, and 3) where the optimal connection points are in the systems for connecting the MVB2B converter. The proposed methodology has been demonstrated by case studies that include various scenarios involving distribution systems with different dominated load types and high photovoltaic penetrations.

## NOMENCLATURE

| | |
|---|---|
| $a$ | Selected bus number for voltage change observation |
| $b$ | Selected bus number for power change implementation |
| $f_1(S_c)$ | Function of converter size for Feeder 1 |
| $f_2(S_c)$ | Function of converter size for Feeder 2 |
| $E_{save}^1$ | Annual energy savings for Feeder 1 |
| $E_{save}^{1-max}$ | Maximum annual energy savings for Feeder 1 |
| $E_{save}^2$ | Annual energy savings for Feeder 2 |
| $E_{save}^{2-max}$ | Maximum annual energy savings for Feeder 2 |
| $l$ | Bus for converter connection |
| $m_1$ | Number of converter sizes for Feeder 1 subset |
| $m_2$ | Number of converter sizes for Feeder 2 subset |
| $NR$ | Net revenue |
| $NR_1$ | Net revenue for Feeder 1 |
| $NR_1^{max}$ | Maximum net revenue for Feeder 1 |
| $NR_2$ | Net revenue for Feeder 2 |
| $NR_2^{max}$ | Maximum net revenue for Feeder 2 |
| $\delta NR_1$ | First derivative of net revenue for Feeder 1 |
| $\delta\delta NR_1$ | Second derivative of net revenue for Feeder 2 |
| $\delta\delta\delta NR_1$ | Third derivative of net revenue for Feeder 3 |
| $n_t$ | Number of total time steps |
| $n_1$ | Number of size options for Feeder 1 |
| $n_2$ | Number of size options for Feeder 2 |
| $n_{yr}$ | Number of years |
| $P_C^{limit}$ | Converter power transfer limit |
| $P_C^{12}$ | Converter power transfer from Feeder 1 to Feeder 2 |
| $P_C^{21}$ | Converter power transfer from Feeder 2 to Feeder 1 |
| $P_{DER}^1(t)$ | DER generation power of Feeder 1 |
| $P_{DER}^2(t)$ | DER generation power of Feeder 2 |
| $P_{limit}^1$ | DER power back-feeding limit for Feeder 1 |
| $P_{limit}^2$ | DER power back-feeding limit for Feeder 2 |
| $P_{load}^1(t)$ | Power consumption of load on Feeder 1 |
| $P_{load}^2(t)$ | Power consumption of load on Feeder 2 |
| $P_{net}^1(t)$ | Net load on Feeder 1 |
| $P_{net}^{1'}(t)$ | Net load on Feeder 1 after power transfer |
| $P_{net}^2(t)$ | Net load on Feeder 2 |
| $P_{net}^{2'}(t)$ | Net load on Feeder 2 after power transfer |
| $P_{save}^1(t)$ | Power saved by converter for Feeder 1 |
| $P_{save}^2(t)$ | Power saved by converter for Feeder 2 |
| $\Delta P$ | Real power change |
| $p_1^{limit}$ | Minimum portion of value retainment for Feeder 1 |
| $p_2^{limit}$ | Minimum portion of value retainment for Feeder 2 |
| $\Delta Q$ | Reactive power change |
| $Ratio_{V\_T}^1$ | Ratio of the value of return to the time of return for Feeder 1 |
| $S_c^i$ | Converter size |
| $S_c^{min}$ | Minimum converter size |
| $S_c^{max}$ | Maximum converter size |
| $S_c^{opt}$ | Optimal converter size |
| $S_{c-1}^{opt-j}$ | Feeder 1 converter size derived from net revenue function derivatives |
| $S_{c-2}^{opt-j}$ | Feeder 2 converter size derived from net revenue function derivatives |
| $S_{c-1}^{opt-tor}$ | Feeder 1 converter size derived from time of return function |
| $S_{c-2}^{opt-tor}$ | Feeder 2 converter size derived from time of return function |
| $S_c^{1-max}$ | Maximum converter size for Feeder 1 |
| $S_c^{2-max}$ | Maximum converter size for Feeder 2 |
| $S_c^{1-min}$ | Minimum converter size for Feeder 1 |
| $S_c^{2-min}$ | Minimum converter size for Feeder 2 |
| $std$ | Standard deviation |
| $ToR_1$ | Time of return for Feeder 1 |
| $\Delta T$ | Time step |
| $VoL_1$ | Value of return for Feeder 1 |
| $VLSMP$ | Voltage load sensitivity matrix for real power |
| $VLSMQ$ | Voltage load sensitivity matrix for reactive power |
| $\Delta V_a$ | Voltage change at node a |
| $\Delta V\_T_1$ | Difference between value of return and time of return |
| $\alpha$ | Weight coefficient |
| $\beta$ | Weight coefficient |
| $\lambda_c$ | Per-unit capital cost of converter |
| $\lambda_{cm}$ | Per-unit maintenance price for the converter |
| $\lambda_{pv}$ | Per-unit price of solar power |

Xiangqi Zhu, Akanksha Singh, and Barry Mather are with the Power Systems Engineering Center, National Renewable1 Energy Laboratory, Golden, Colorado 80401, USA. (e-mails: xiangqi.zhu@nrel.gov, akanksha.singh@nrel.gov, barry.mather@nrel.gov).

| | |
|---|---|
| $\gamma_1$ | Portion of Feeder 1 energy savings with respect to total energy savings of two feeders |
| $\gamma_2$ | Portion of Feeder 2 energy savings with respect to total energy savings of two feeders |
| $\eta$ | Power transfer efficiency |
| $\mu_1$ | Average load of Feeder 1 |
| $\mu_2$ | Average load of Feeder 2 |

## I. INTRODUCTION

THE variability and uncertainty of distributed energy resources (DERs) have been a major challenge for DER grid integration. The state of the art has a wide range of methods aimed at addressing this challenge, including managing the consumption at the demand side [1]–[3], pairing renewable generation with energy storage [4]–[7], and developing advanced generation dispatch approaches [8]–[10]. These methods all focus on resolving the problems locally within a distribution system; however, if two or more distribution systems can be connected with advanced devices that can effectively control the power exchange between them, the connected systems can complement each other and more efficiently resolve the DER integration challenges.

Recently, the technology of medium-voltage back-to-back (MVB2B) converters, which can connect two distribution systems, has had great improvement, with multiple advanced functions developed [11]. With the function of quantifiable power transfer, the MVB2B converter can effectively transfer power from the feeder with excess DER generation to the feeder with less DER generation. In this way, the DER generation can be effectively used without a substantial portion of curtailment.

Along with the technology advancement of the MVB2B converter, the system-level analysis and development, such as value analysis and implementation strategies, become critical to enable the commercial adoption of MVB2B converters.

Most research in the literature on MVB2B converters focuses on the converter technology, such as the design of the architecture and control [12]–[13], and performance improvement [14]–[15]. The system-level analysis and development has been barely touched. Critical questions on the system-level implementation of the MVB2B converter have not been answered, such as how much benefit the converter can bring to the system and which converter capacity is optimal for different connected distribution systems.

It is crucial to answer these questions to provide grid operators comprehensive information for the system implementation of the MVB2B converter and to provide effective methods of selecting the optimal converter size for connected systems.

Our previous work [16]–[17] investigated, analyzed, and quantified the value the MVB2B converter can bring to connected systems. The analysis we performed for various cases shows that the MVB2B converter can contribute significant value to enhancing distribution system DER hosting.

To enable the MVB2B converter to contribute the most value, it is critical to choose the appropriate distribution systems to be connected and implement an optimal size for the connected systems. As continuous work to our previous value analysis of the MVB2B converter, in this paper, we develop a systematic methodology to address three important challenges in the real-world implementation of the MVB2B converter.

First, to address the challenge of the MVB2B converter sizing, we propose a method to select the optimal size for a pair of systems to be connected, with cost-benefit considerations. Second, to tackle the problem of the system pair selection, we propose a method to evaluate the suitability of two distribution systems to be connected, according to the aggregated load profiles of the systems. Third, we propose a method to locate the optimal connection node on a system to help system operators select the node for connecting the MVB2B converter.

The rest of the paper is organized as follows: Section II briefly introduces the MVB2B converter and its functions. Section III presents the methodology we developed for preparing the converter for real-world implementation, including converter sizing, system selection, and connection point selection. Section IV discusses the case studies, followed by Section V, which concludes the paper.

## II. MVB2B CONVERTER INTRODUCTION

This section briefly presents the topology and the control loops of the MVB2B converter. In particular, the control function that governs the quantifiable power transfer function, which provides great value to the DER hosting enhancement, is described here.

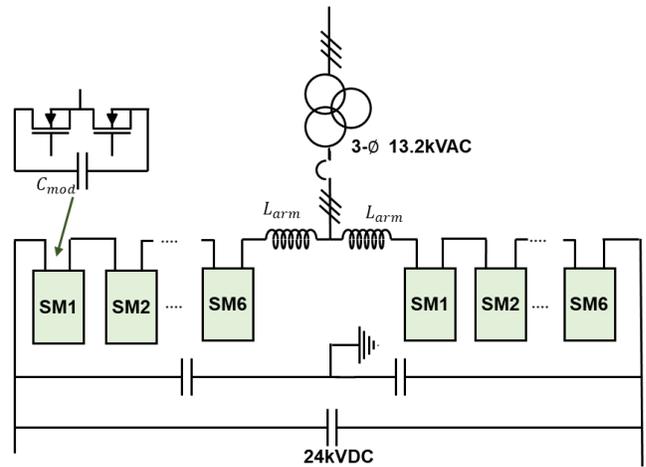

Fig. 1 Topology of one side of the MVB2B converter

A double-star chopper-cell (DSCC) topology [18] is selected to implement the back-to-back connected, modular, multilevel converters. The architecture of one side of the MVB2B converter is shown in Fig. 1. Here, *SM* denotes the submodules, which consist of a half-bridge leg, with each switch rated at 10 kV. To control the DSCC converter, two forms of switching-level control-voltage/current controls are implemented [19]. The application of the controls is in the form of an averaged current control setup, with the voltage loop being the outer one and the current loop being the inner one. The voltage control is further split into three different controllers: active power, reactive power, and dc-link control. Specifically, the following control is implemented to realize the quantifiable power

transfer function：

The controller is based on the transformation of the three-phase grid signals to the direct-quadrature (dq) frame using the conventional Park/Clark theories. Mathematical expressions of the same can be sourced from [18]. In doing this transformation, we deal with only 2 components, instead of 3, which reduces the mathematical complexities. The direct axis voltage is subtracted from the reference provided by the user and compensated using a proportional-integral controller. This is used as the active power controller. A similar form of control is implemented for the reactive control loop, where the quadrature axis voltage signal is used. The architecture and operation of the controllers are described in [20]. Additionally, advanced grid support functions are implemented as upper controls. These grid support functions generate the reference active power, reactive power, and dc-link voltage based on the grid conditions monitored through sensed voltage and frequency. The details of this implementation and evaluation are presented in **Error! Reference source not found.**.

### III. Methodology

This section presents three subsections that introduce, respectively, the proposed methodologies for the size selection, the feeder pair selection, and the connection point selection.

The size selection approach helps utilities choose an optimal MVB2B converter size for the two distribution systems connected through it. The feeder pair selection approach helps choose two systems that mutually benefit if they are connected with the MVB2B converter.

The connection point selection approach helps choose the appropriate nodes on the two distribution systems to connect the MVB2B converter with minimal system impact. The selected connection points will be able to receive or output a substantial amount of power without affecting the system voltage or causing power congestion.

#### A. Size Selection

The equations shown in (1)–(11) constitute the model of the power transfer through the MVB2B converter between the two connected feeders. Generally, when there is excess DER generation on one feeder and there is load not covered by the DER generation on the other feeder, the excess power that is under the converter limit will be transferred to the other feeder. The DER power savings from the curtailment reduction brought by the MVB2B converter connection are calculated in (7) and (11) for the two feeders, respectively.

The total DER energy savings in the whole simulation period is calculated in (12) and (13). Here, the simulation length is set to be 1 year. Then the total net revenue the converter can bring to the two feeders is calculated in (14). The net revenue for the two feeders separately is calculated in (15) and (16), respectively. The capital and maintenance costs that the two feeders need to undertake is weighted using the coefficients calculated in (17) and (18). The feeder that can obtain more energy savings from the MVB2B converter will pay a larger portion of capital and maintenance costs.

$$P_{net}^1(t) = P_{load}^1(t) - P_{DER}^1(t) \tag{1}$$

$$P_{net}^2(t) = P_{load}^2(t) - P_{DER}^2(t) \tag{2}$$

$$P_C^{limit} = \eta S_c \tag{3}$$

When $P_{net}^1(t) < 0, \ P_{net}^2(t) > 0$

$$P_C^{12}(t) = \min\left(|P_{net}^1(t)|, P_{net}^2(t), P_C^{limit}\right) \tag{4}$$

$$P_{net}^{1\prime}(t) = P_{net}^1(t) + P_C^{12}(t) \tag{5}$$

$$P_{net}^{2\prime}(t) = P_{net}^2(t) - P_C^{12}(t) \tag{6}$$

$$P_{save}^1(t) = P_C^{12}(t) - P_{limit}^1 \tag{7}$$

When $P_{net}^1(t) > 0, \ P_{net}^2(t) < 0,$

$$P_C^{21}(t) = \min\left(P_{net}^1(t), |P_{net}^2(t)|, , P_C^{limit}\right) \tag{8}$$

$$P_{net}^{1\prime}(t) = P_{net}^1(t) - P_C^{21}(t) \tag{9}$$

$$P_{net}^{2\prime}(t) = P_{net}^2(t) + P_C^{21}(t) \tag{10}$$

$$P_{save}^2(t) = P_C^{21}(t) - P_{limit}^2 \tag{11}$$

$$E_{save}^1 = \sum_{t=1}^{n_t} P_{save}^1(t) \cdot \Delta T \tag{12}$$

$$E_{save}^2 = \sum_{t=1}^{n_t} P_{save}^2(t) \cdot \Delta T \tag{13}$$

$$NR = n_{yr}\lambda_{pv}(E_{save}^1 + E_{save}^2) - \lambda_c S_c - n_{yr}\lambda_{cm} \tag{14}$$

$$NR_1 = n_{yr}\lambda_{pv}(E_{save}^1) - \gamma_1(\lambda_c S_c + n_{yr}\lambda_{cm}) \tag{15}$$

$$NR_2 = n_{yr}\lambda_{pv}(E_{save}^2) - \gamma_2(\lambda_c S_c + n_{yr}\lambda_{cm}) \tag{16}$$

$$\gamma_1 = \frac{E_{save}^{1-max}}{E_{save}^{1-max} + E_{save}^{2-max}} \tag{17}$$

$$\gamma_2 = \frac{E_{save}^{2-max}}{E_{save}^{1-max} + E_{save}^{2-max}} \tag{18}$$

As shown in (19)–(23), the net revenue is a function of converter size. Here, $n_{yr}\lambda_{pv}(f_1(S_c) + f_1(S_c))$ represents the revenue of the DER energy savings from curtailment, $\lambda_c S_c$ represents the converter capital cost, and $n_{yr}\lambda_{cm}$ is the maintenance cost.

$$E_{save}^1 = f_1(S_c) \tag{19}$$

$$E_{save}^2 = f_2(S_c) \tag{20}$$

$$NR(S_c) = n_{yr}\lambda_{pv}(f_1(S_c) + f_1(S_c)) - \lambda_c S_c - n_{yr}\lambda_{cm} \tag{21}$$

$$NR_1(S_c) = n_{yr}\lambda_{pv}(f_1(S_c)) - \gamma_1(\lambda_c S_c - n_{yr}\lambda_{cm}) \tag{22}$$

$$NR_2(S_c) = n_{yr}\lambda_{pv}(f_2(S_c)) - \gamma_2(\lambda_c S_c - n_{yr}\lambda_{cm}) \tag{23}$$

When the size of the converter reaches a certain point, the maximum net revenue will be achieved for the two feeders, as shown in (24) and (25). The net revenue of the feeder will start decreasing if the converter size keeps increasing after the maximum.

$$NR_1^{max} = n_{yr}\lambda_{pv}(f_1(S_c^{1-max})) - \gamma_1(\lambda_c S_c^{1-max} - n_{yr}\lambda_{cm}) \quad (24)$$

$$NR_2^{max} = n_{yr}\lambda_{pv}(f_1(S_c^{2-max})) - \gamma_2(\lambda_c S_c^{2-max} - n_{yr}\lambda_{cm}) \quad (25)$$

The size selection based on the net revenue functions in (24)–(25) is introduced in the following part. Here, we take Feeder 1 as an example, which means that the following formulations (26)–(28) are based on the net revenue function for Feeder 1.

The first-order derivative of the net revenue function in (24) is calculated in (26). The first-order derivative also represents the marginal value the converter can bring to the feeder by increasing one unit of the converter size.

To measure the speed of the decreasing marginal value and to capture the speed change, the second- and third-order derivative of (24) have also been calculated, as shown in (27)–(28).

$$\delta NR_1(S_c^i) = NR_1(S_c^i) - NR_1(S_c^{i-1}) \quad (26)$$

$$\delta\delta NR_1(S_c^i) = \delta NR_1(S_c^i) - \delta NR_1(S_c^{i-1}) \quad (27)$$

$$\delta\delta\delta NR_1(S_c^{i+2}) = \delta\delta NR_1(S_c^{i+2}) - \delta\delta NR_1(S_c^{i+1}) \quad (28)$$

In addition to the net revenue, the time of return is an important criterion to select the optimal converter size. The time needed to pay back the capital cost is calculated in (29)–30). An optimal converter size should have better value and less time of return. To simultaneously compare the total value the converter can bring and the time needed to pay back the capital cost, as shown in (31)–(32), we use a ratio to properly scale down the total value and make its magnitude comparable to the time of return. In this way, we can calculate the difference between the total value and the time of return to determine the optimal size that can bring higher value and has less time of return. The method of using these three derivatives and time of return to select the optimal converter size are discussed in the following.

$$ToR_1(S_c) = \frac{\lambda_c S_c}{\lambda_{pv}(f_1(S_c))} \quad (29)$$

$$VoL_1(S_c) = n_{yr}\lambda_{pv}(f_1(S_c)) \quad (30)$$

$$Ratio_{V\_T}^1 = \frac{max(\ |VoL_1(S_c^i)|_{1\times n})}{max(|ToR_1(S_c^i)|_{1\times n})} \quad (31)$$

$$\Delta V\_T_1 = \frac{VoL_1(S_c)}{Ratio_{VT}^1} - ToR_1(S_c) \quad (32)$$

When the first-order derivative reaches zero, we have the maximum size we can select, as shown in (33). As shown in (34)–(35), the sizes where the monotonicity of the second or third derivative changes are critical sizes, which are the candidates for the optimal size. Here, we take Feeder 1 as an example; the calculation for Feeder 2 is the same.

$$\text{When} \quad \delta NR_1(S_c^i)=0, \quad S_{c-1}^{max} = S_c^i \quad (33)$$

When $\delta\delta NR_1(S_c^i) < \delta\delta NR_1(S_c^{i+1})$ and $\delta\delta NR_1(S_c^i) < \delta\delta NR_1(S_c^{i-1})$
or $\delta\delta NR_1(S_c^i) > \delta\delta NR_1(S_c^{i+1})$ and $\delta\delta NR_1(S_c^i) > \delta\delta NR_1(S_c^{i-1})$

$$S_{c-1}^{opt-j} = S_c^i \quad (34)$$

When $\delta\delta\delta NR_1(S_c^i) < \delta\delta\delta NR_1(S_c^{i+1})$ and $\delta\delta\delta NR_1(S_c^i) < \delta\delta\delta NR_1(S_c^{i-1})$
or $\delta\delta\delta NR_1(S_c^i) > \delta\delta\delta NR_1(S_c^{i+1})$ and $\delta\delta\delta NR_1(S_c^i) > \delta\delta\delta NR_1(S_c^{i-1})$

$$S_{c-1}^{opt-j} = S_c^i \quad (35)$$

The optimal size candidate for Feeder1, which is selected by comparing the total value and the time of return, is the size when $\Delta V\_T_1$ in (32) is maximized, as shown in (36). The candidate for Feeder 2 can be selected in the same way.

$$S_{c-1}^{opt-tor} = S_c^i \quad \text{when} \quad \Delta V\_T \text{ reaches max} \quad (36)$$

To guarantee that retaining a substantial portion of the maximum value the converter can bring, as shown in (37)–(40), a limit will be set for each feeder, respectively. The final selected optimal size needs to bring a value that has a portion larger than the minimum limit.

$$p_1^{limit} = \frac{NR_1(S_{c-1}^{min})}{NR_1^{max}} \quad (37)$$

$$p_2^{limit} = \frac{NR_2(S_{c-2}^{min})}{NR_2^{max}} \quad (38)$$

Based on the limit set, we will have a minimum size for the two feeders, $S_{c-1}^{min}$ and $S_{c-2}^{min}$. As shown in (39), the minimum size for these two feeders is calculated by taking the larger one between them.

Similarly, the maximum converter size for the two feeders is calculated in (40).

$$S_c^{min} = \max(S_{c-1}^{min}, S_{c-2}^{min}) \quad (39)$$

$$S_c^{max} = \max(S_{c-1}^{max}, S_{c-2}^{max}) \quad (40)$$

Now, we have a list of size options for the two feeders, as shown in Table I.

TABLE I. SIZE OPTIONS

| Option number | 1 | 2 | 3 | 4 | 5 | 6 |
|---|---|---|---|---|---|---|
| Size | $S_c^{min}$ | $\|S_{c-1}^{opt-j}\|_{1\times n_1}$ | $\|S_{c-2}^{opt-j}\|_{1\times n_2}$ | $S_{c-1}^{opt-tor}$ | $S_{c-2}^{opt-tor}$ | $S_c^{max}$ |

If all the sizes in 2, 3, 4, or 5 are smaller than $S_c^{min}$, then:

$$S_c^{opt} = S_c^{min} \quad (41)$$

But if there are sizes in 2, 3, 4, and 5 bigger than $S_c^{min}$, then a subset of sizes will be determined where every option is larger than $S_c^{min}$ and smaller than $S_c^{max}$

TABLE II. SIZE OPTION SUBSET

| Option number | 1 | 2 | 3 | 4 |
|---|---|---|---|---|
| Size | $\left|S_{c-1}^{opt-j}\right|_{1\times m_1}$ | $\left|S_{c-2}^{opt-j}\right|_{1\times m_2}$ | $S_{c-1}^{opt-tor}$ | $S_{c-2}^{opt-tor}$ |

Then, as shown in (42), the optimal size for the two connected distribution systems is calculated by taking the larger value in each option and selecting the smallest one from the larger numbers. In this way, we can select a converter size that achieves a balance between securing a substantial amount of energy savings value and being paid back without waiting for a long time.

$$S_c^{opt} = \min\left(\max\left(\left|S_{c-1}^{opt-t}\right|_{1\times m_1}\right), \max\left(\left|S_{c-2}^{opt-t}\right|_{1\times m_2}\right), \max(S_{c-1}^{opt-tor}, S_{c-2}^{opt-tor})\right) \quad (42)$$

### B. Feeder Pair Selection

Within the geographic allowance, it is important to connect two feeders that can yield substantial value on DER enhancement. In this way, the value of the MVB2B converter can be maximized, and the years of return can be shorter for the two feeders that adopted the MVB2B converter.

Based on the extensive simulations we have conducted in our value analysis work, an empirical conclusion can be drawn, as follows: The best value comes from the pairs where one feeder has a higher peak load and is dominated with commercial load and the other feeder has a lower peak load and is dominated by residential load.

Developed from our extensive simulations, as shown in (43)–(45), we propose using the standard deviation of the feeder head load profile as a criterion to select the feeder pairs. The summation of the standard deviation of the load profiles measured at the substations of the two feeders can effectively indicate the value that the two feeders can obtain from implementing the MVB2B converter. Generally, two feeders with load profiles featuring higher standard deviation values will gain higher value on DER hosting enhancement through the MVB2B converter and can reach a cost-benefit balance within a shorter time.

$$std = \sqrt{\frac{1}{n_t}\sum_{t=1}^{n_t}(P_{load}^1(t) - \mu_1)} + \sqrt{\frac{1}{n_t}\sum_{t=1}^{n_t}(P_{load}^2(t) - \mu_2)} \quad (43)$$

$$\mu_1 = \frac{1}{n_t}\sum_{t=1}^{n_t} P_{load}^1(t) \quad (44)$$

$$\mu_2 = \frac{1}{n_t}\sum_{t=1}^{n_t} P_{load}^2(t) \quad (45)$$

### C. Connection Point Selection

We propose to leverage the voltage load sensitivity matrix (VLSM) developed in [20] to help select the converter connection point on the feeder. As shown in (46)–(47), the VLSM can be developed for real and reactive power, respectively, and represents the voltage sensitivity of one bus to the power change at other buses. The voltage change at bus $a$ that is caused by the power change at bus $b$ can be calculated using (48), which is derived from (47) and is formulated based on the real/reactive power sensitivity factors.

Considering that the converter transfers real power between two feeders, the voltage change at bus $a$ caused by the converter power transfer at connection bus $l$ can be calculated by (49). The total voltage changes happening at all the other buses can be calculated by (50). Because the power transfer amount can be considered a constant when comparing the total voltage changes caused by different connection points, $l$, the comparison of voltage changes can be simplified as the comparison of the summation of the voltage sensitivity factors, as shown in (51).

Because the power transfer through the converter can be a large amount, it is important to maintain the stable system voltage when the converter transfers power between the two distribution systems; therefore, we want to minimize the $p_{sum}$ in (51). In addition to the voltage stability, in real-world implementation, we need to consider the distance between the connection point and the major DERs and balance the distance with the voltage changes, as shown in (52). Here, the coefficient $r$ is for adjusting the magnitude of the distance summation to a comparison level of the sensitivity factor summation.

If there are large-capacity DERs, such as solar and wind power plants, it is better to set the weight coefficient, $\beta$, at a higher value so that the excess power generated by the solar/wind plants can be transferred to another distribution system within a minimum length of line and impact a minimum part of the system. If all the DERs are fairly distributed along the distribution system, the weight coefficient, $\beta$, can be set at a lower value because the distance between the connection point and the DERs has much less impact on the system in this case.

$$|\Delta V| = |VLSMP||\Delta P| + |VLSMQ||\Delta Q| \quad (46)$$

i.e.,

$$\begin{vmatrix}\Delta V_1\\ \vdots\\ \Delta V_{n_{node}}\end{vmatrix} = \begin{vmatrix}p_{11} & \cdots & p_{1n_{node}}\\ \vdots & \ddots & \vdots\\ p_{n1} & \cdots & p_{n_{node}n_{node}}\end{vmatrix}\begin{vmatrix}\Delta P_1\\ \vdots\\ \Delta P_{n_{node}}\end{vmatrix} + \begin{vmatrix}q_{11} & \cdots & q_{1n_{node}}\\ \vdots & \ddots & \vdots\\ q_{n_{node}1} & \cdots & q_{n_{node}n_{node}}\end{vmatrix}\begin{vmatrix}\Delta Q_1\\ \vdots\\ \Delta Q_{n_{node}}\end{vmatrix} \quad (47)$$

$$\Delta V_a = \sum_{b=1}^{n_{node}} p_{ab}\Delta P_b + \sum_{b=1}^{n_{node}} q_{ab}\Delta Q_b \quad (48)$$

$$\Delta V_a = VLSMP(a,l)\Delta P_l \quad (49)$$

$$\Delta V_{sum} = \sum_{a=1}^{n_{node}} VLSMP(a,l)\Delta P_l \quad (50)$$

$$p_{sum} = \sum_{a=1}^{n_{node}} VLSMP(a,l) \quad (51)$$

$$\text{Min } C = \alpha\sum_{a=1}^{n_{node}} VLSMP(a,l) + \beta \cdot r\sum_{b=1}^{n_{DER}} d(b,l) \quad (52)$$

$$\alpha + \beta = 1 \quad (53)$$

## IV. CASE STUDY

In this section, we present and analyze the case studies that demonstrate the proposed methodology for the converter sizing,

the feeder pair selection, and the connection point selection.

*A. Case Studies on Converter Size Selection*

We selected two case studies for the MVB2B converter sizing. In the first case, Feeder 1 is dominated by residential load, whereas Feeder 2 is dominated by commercial load, and the peak load of Feeder 1 is approximately 80% of the peak load of Feeder 2. Both feeders have 100% photovoltaic (PV) penetration. The load profiles of the two feeders are all aggregated from realistic field-measured data of residential houses and commercial buildings, with 1-year length and 30-minute resolution.

Fig. 2 presents the yearly PV energy savings trend and the 10-year accumulated net revenue trend, along with the converter capacity increasing from 200 kva to 1500 kva. Here, we assume that the PV energy price, $\lambda_{pv}$, is \$0.1/kWh and the per-unit capital cost of the converter, $\lambda_c$, is \$100/kva. In real-world calculation, actual prices can be implemented to replace those assumed prices. We can see that 900 kva is the size that can reach the maximum PV energy savings for Feeder 1 if the cost-benefit is not considered. But if we use net revenue as a criterion, the optimal size of the converter for Feeder 1 is reduced to 700 kva.

To better articulate the optimal size selection, we replot the net revenue trends, $NR_1(S_c)$ and $NR_2(S_c)$, in Fig. 3. As indicated by the two black horizontal lines in Fig. 3 and shown in (54), the maximum sizes for Feeder 1 and Feeder 2 are 700 kva and 450 kva, respectively. Then the maximum size we set for the two feeders is 700 kva.

$$S_c^{max} = \max(S_{c-1}^{max}, S_{c-2}^{max})$$
$$= \max(700\ kva, 450\ kva) = 700 kva \quad (54)$$

If we set the limit as 80% for both $p_1^{limit}$ and $p_2^{limit}$, as indicated by the two green horizontal lines, then the minimum sizes can be obtained by analyzing Fig. 3. The results are shown in (55).

$$S_c^{min} = \max(S_{c-1}^{min}, S_{c-2}^{min})$$
$$= \max(350\ kva, 250\ kva) = 350\ kva \quad (55)$$

As presented in Section III, by analyzing the turning points in the second and the third derivatives of $NR_1(S_c)$ and $NR_2(S_c)$, which are shown in Fig. 4 (a) and Fig. 4 (b), respectively, we can obtain the size values for $|S_{c-1}^{opt-j}|_{1\times n_1}$ and $|S_{c-2}^{opt-j}|_{1\times n_2}$, which are shown in Table III.

Fig. 5 shows the time of return and the value it can bring for each converter capacity. As indicated in Fig. 5, the converter sizes that can bring better value and less time of return for feeders 1 and 2 are 600 kva and 350 kva, respectively. Those two values are also shown in Table III.

Then a subset of the size options can be derived in Table IV. By bringing the values in Table IV into (42), presented in Section III, we can derive an appropriate size for connecting the two feeders, as shown in (56), which is 400 kva.

TABLE III. SIZE OPTIONS

| Option number | 1 | 2 | 3 | 4 | 5 | 6 |
|---|---|---|---|---|---|---|
| Size | $S_c^{min}$ | $\|S_{c-1}^{opt-j}\|_{1\times n_1}$ | $\|S_{c-2}^{opt-j}\|_{1\times n_2}$ | $S_{c-1}^{opt-tor}$ | $S_{c-2}^{opt-tor}$ | $S_c^{max}$ |
| Value (kva) | 350 | $\|500\|_{1\times 1}$ | $\|400,450\|_{1\times 2}$ | 600 | 350 | 700 |

TABLE IV. SIZE OPTION SUBSET

| Option number | 1 | 2 | 3 | 4 |
|---|---|---|---|---|
| Size | $\|S_{c-1}^{opt-j}\|_{1\times m_1}$ | $\|S_{c-2}^{opt-j}\|_{1\times m_2}$ | $S_{c-1}^{opt-tor}$ | $S_{c-2}^{opt-tor}$ |
| Value | 500 | 400 | 600 | 350 |

$$S_c^{opt} = \min\left(\max\left(|S_{c-1}^{opt-t}|_{1\times m_1}\right), \max\left(|S_{c-2}^{opt-t}|_{1\times m_2}\right), \max(S_{c-1}^{opt-tor}, S_{c-2}^{opt-tor})\right)$$
$$= \min(500\ kva, 400\ kva, 600\ kva) = 400 kva \quad (56)$$

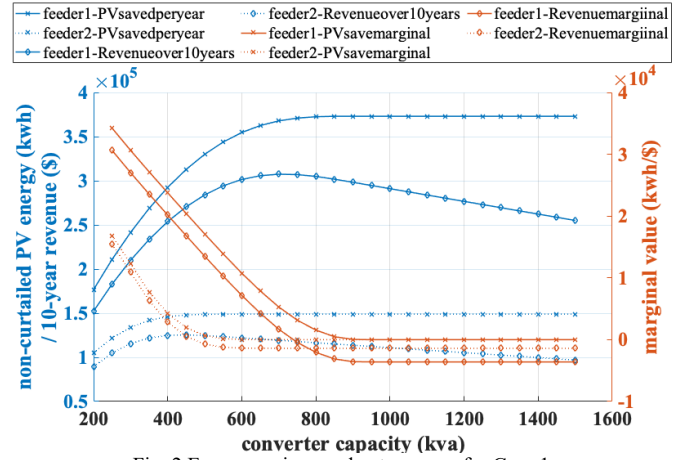
Fig. 2 Energy savings and net revenue for Case 1

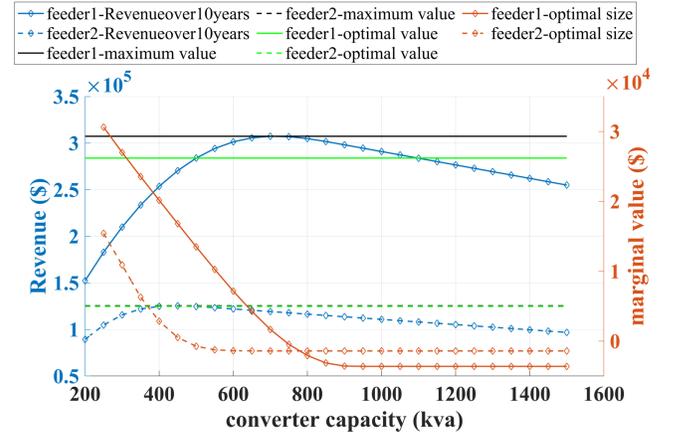
Fig. 3 Net revenue for Case 1

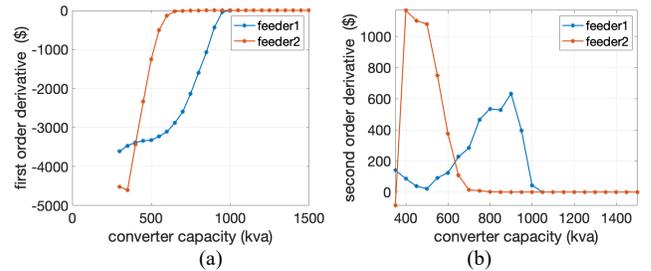
Fig. 4 Second derivative (a) and third derivative (b)

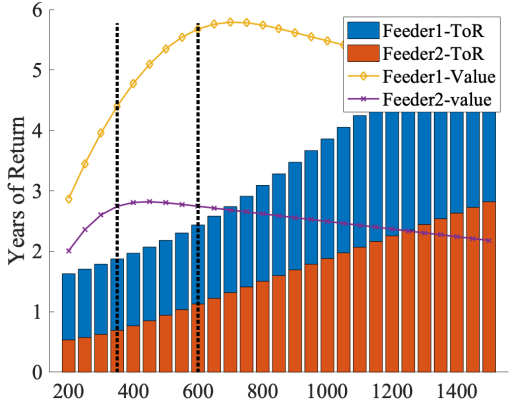

Fig. 5 Total value over 10 years versus time of return for Case 1

Figs. 6–9 demonstrate the converter sizing analysis for another case. In this case, feeders 1 and 2 are dominated by residential and commercial load, respectively, where the peak load of Feeder 1 is approximately 50% of the peak load of Feeder 2.

Similar to Case 1, Fig. 6 demonstrates the maximum size difference before and after adding the cost-benefit to the considerations. Here, the maximum size for the maximum PV energy savings of Feeder 1 is 1150 kva, whereas the maximum size considering the net revenue is 900 kva.

Using the size selection methodology presented in Part A of Section III, we can derive the size values shown in Table V from Figs. 7–9.

From Fig. 7, we can derive the maximum sizes for Feeder 1 and 2 as 900 kva and 500 kva, respectively. From Fig. 8, we can derive the sizes that are turning points for each feeder, as shown in Table V. As show in (57) and (58), the maximum and minimum converter sizes for connecting the two feeders are 900 kva and 550 kva, respectively. The critical sizes that maximize the value while minimizing the time of return for the two feeders can be derived from Fig. 9.

Here, when we derive a subset, shown in Table VI, from Table V, because the minimum size is 550 kva, then all the sizes smaller than 550 kva have been discarded. The final selected optimal size is calculated in (59).

$$S_c^{max} = \max(S_{c-1}^{max}, S_{c-2}^{max})$$
$$= \max(900\ kva, 550\ kva) = 900\ kva \quad (57)$$

If we set the limit as 80% for both $p_1^{limit}$ and $p_2^{limit}$, then:

$$S_c^{min} = \max(S_{c-1}^{min}, S_{c-2}^{min})$$
$$= \max(550\ kva, 300\ kva) = 550\ kva \quad (58)$$

TABLE V. SIZE OPTIONS

| Option number | 1 | 2 | 3 | 4 | 5 | 6 |
|---|---|---|---|---|---|---|
| Size | $S_c^{min}$ | $\lvert S_{c-1}^{opt-j}\rvert_{1\times n_1}$ | $\lvert S_{c-2}^{opt-j}\rvert_{1\times n_2}$ | $S_{c-1}^{opt-tor}$ | $S_{c-2}^{opt-tor}$ | $S_c^{max}$ |
| Value (kva) | 550 | $\lvert 400, 600, 700\rvert_{1\times 3}$ | $\lvert 400\rvert_{1\times 1}$ | 700 | 450 | 900 |

TABLE IV. SIZE OPTION SUBSET

| Option number | 1 | 3 |
|---|---|---|

| Size | $\lvert S_{c-1}^{opt-j}\rvert_{1\times m_1}$ | $S_{c-1}^{opt-tor}$ |
|---|---|---|
| Value | $\lvert 600, 700\rvert_{1\times 2}$ | 700 |

$$S_c^{opt} = \min\left(\max\left(\lvert S_{c-1}^{opt-t}\rvert_{1\times m_1}\right), \max\left(\lvert S_{c-2}^{opt-t}\rvert_{1\times m_2}\right), \max(S_{c-1}^{opt-tor}, S_{c-2}^{opt-tor})\right)$$
$$= \min(700\ kva, 700\ kva) = 700\ kva \quad (59)$$

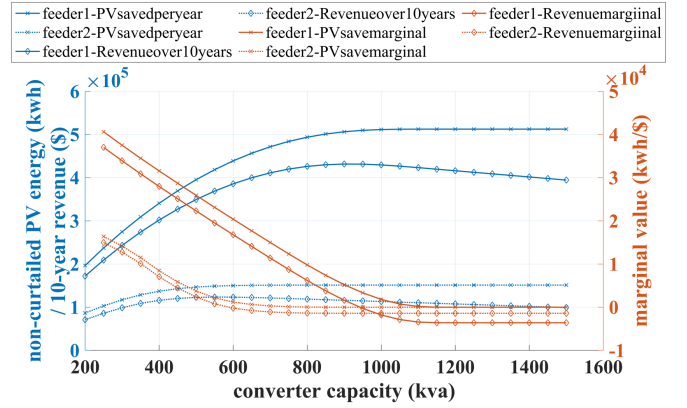

Fig. 6 Energy savings and net revenue for Case 2

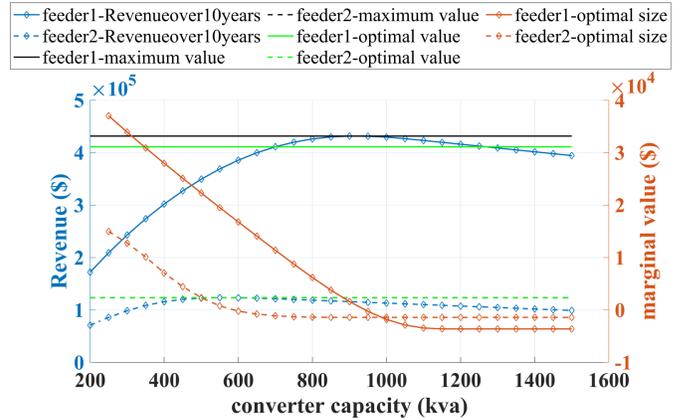

Fig. 7 Net revenue for Case 1

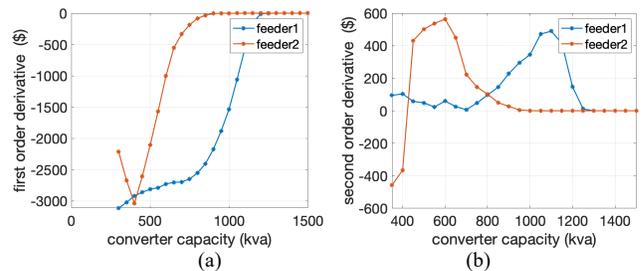

Fig. 8 Second derivative (a) and third derivative (b)

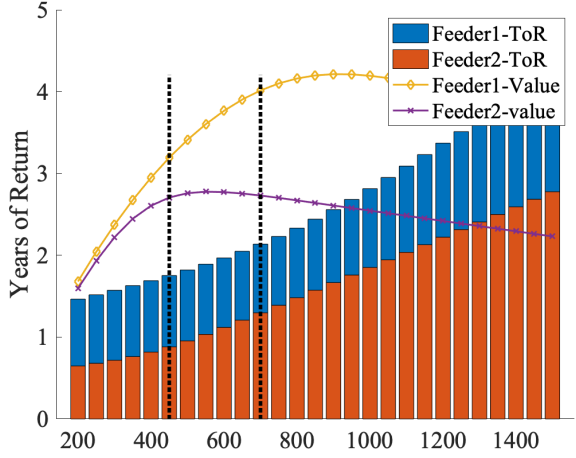

Fig. 9 Total value over 10 years versus time of return for Case 2

### B. Case Study on Feeder Pair Selection

To demonstrate the feeder pair selection method presented in Part B in Section III, we calculated the yearly PV energy savings and the standard deviation summation of the feeder load profiles for 24 scenarios.

As shown in Fig. 10, the 24 scenarios have been divided into 4 areas, where each area includes 6 scenarios. From left to right, the first area covers the scenarios where both feeders are dominated by commercial load, the second area includes scenarios where both feeders are dominated by residential load, the third and fourth areas cover the scenarios where one feeder is dominated by residential load and the other feeder is dominated by residential load. In the third area, the commercial feeder has a higher peak, whereas the residential feeder has a higher peak in the fourth area.

Here, we define a peak load ratio as shown in (60). In each area, from left to right, the $Pratio$ is discretely increasing from 30% to 80%, with 10% as a step.

$$Pratio = \frac{\max(P_{load}^1(t))}{\max(P_{load}^2(t))} \quad (60)$$

As shown in Fig. 10, the yearly PV energy savings and the standard deviation summation have very similar trends, along with the scenarios, which demonstrates that we can use the standard deviation of the feeder head load profile to effectively select the feeder pairs.

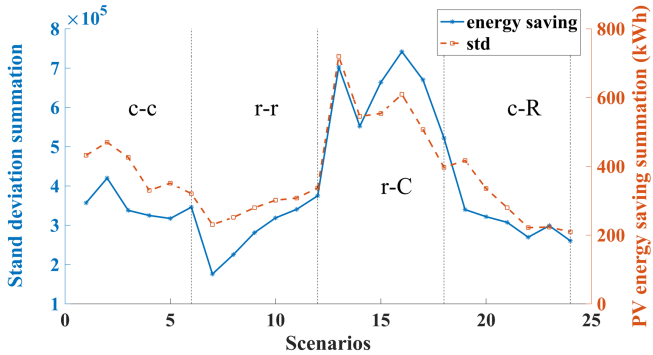

Fig. 10 The energy savings and standard deviation trends

### C. Case Study on Connection Point Selection

As shown in Fig. 11, we placed three solar plants at a realistic utility feeder model and demonstrate the connection point selection for this feeder scenario.

Here, we assume $\alpha$ and $\beta$ both have a value of 0.5, which means we weight the distance to the DERs and the voltage sensitivity as equally important. Fig. 12 shows the summation of distance to the three solar plants from each node and the voltage sensitivity of each node, which are the second and the first part of (52), shown in Part C of Section III. Fig. 12 shows that the first couple nodes that are close to feeder head have the lowest voltage sensitivity value, whereas the nodes in the middle of the feeder have the shortest distance to the solar plants.

Fig. 13 shows the C value of each node, which is calculated using (52). The node with the lowest C value is the selected connection point, which is node 142, as indicated using a red line in Fig. 13 and a green circle in Fig. 11.

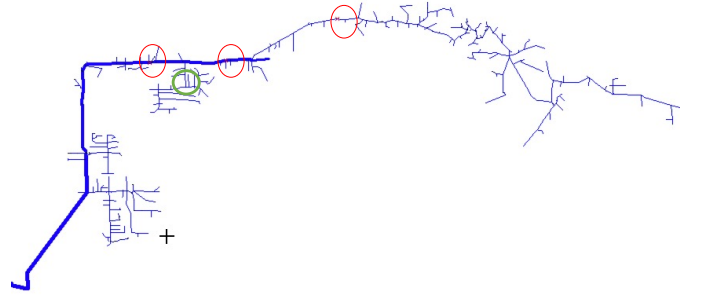

Fig. 11 Feeder model topology

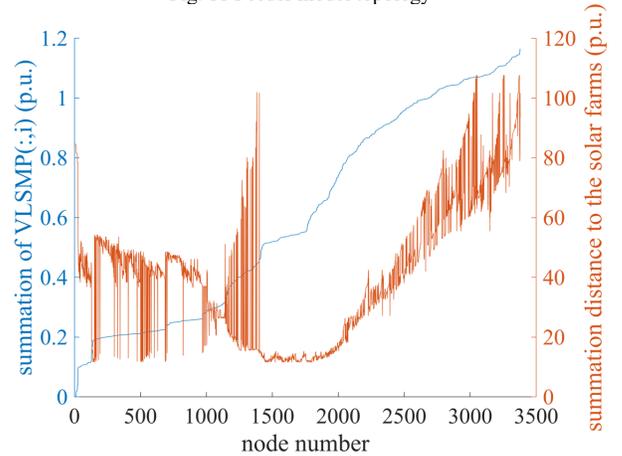

Fig. 12 voltage sensitivity and distance to solar plants of the nodes

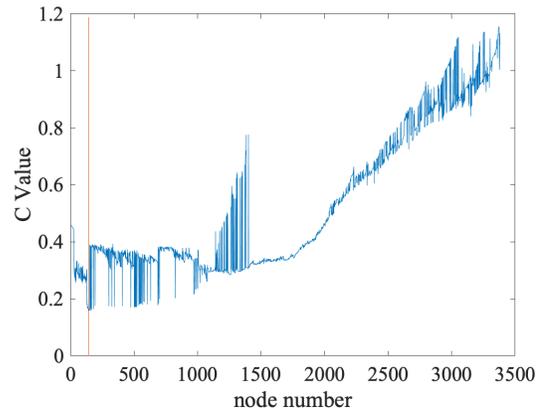

Fig. 13 C value of the nodes

## V. Conclusion

In this work, we developed a methodology to enable MVB2B converter real-world implementation by addressing three critical problems: 1) selecting the optimal converter size for connecting two distribution systems, 2) selecting the appropriate systems to be connected that can mutually benefit, and 3) selecting the system node that is suitable for connecting the converter. We develop the converter sizing method based on the cost-benefit function and time-of-return function. By extracting the prominent load profile feature, we can successfully select the systems that can mutually benefit if they are connected by the MVB2B converter. Then we leverage the VLSM in [20] to select the appropriate nodes in the systems to connect the MVB2B converter. The proposed methods are all demonstrated in our case study.

## Acknowledgments

This work was authored by Alliance for Sustainable Energy, LLC, the manager and operator of the National Renewable Energy Laboratory for the U.S. Department of Energy (DOE) under Contract No. DE-AC36-08GO28308. Funding provided by U.S. Department of Energy Office of Energy Efficiency and Renewable Energy Advanced Manufacturing Office via the GADTAMS project. The views expressed in the article do not necessarily represent the views of the DOE or the U.S. Government. The U.S. Government retains and the publisher, by accepting the article for publication, acknowledges that the U.S. Government retains a nonexclusive, paid-up, irrevocable, worldwide license to publish or reproduce the published form of this work, or allow others to do so, for U.S. Government purposes.